\documentclass[12pt]{article}

\usepackage[centertags]{amsmath}
\usepackage{amsfonts}
\usepackage{amssymb}
\usepackage{amsthm}
\usepackage{newlfont}
\usepackage{epsfig}
\usepackage{amscd}

\newcommand{\ZZ}{{\mathbb Z}}

\newcommand{\beq}{\begin{equation}}
\newcommand{\eeq}{\end{equation}}
\newcommand{\ba}{\begin{array}}
\newcommand{\ea}{\end{array}}
\newcommand{\bea}{\begin{eqnarray}}
\newcommand{\eea}{\end{eqnarray}}
\newcommand{\Di}{\Delta_1}
\newcommand{\Dj}{\Delta_2} 
\newcommand{\Dim}{\Delta_{-1}}
\newcommand{\Djm}{\Delta_{-2}}

\newcommand{\I}{_{1}}
\newcommand{\J}{_{2}}
\newcommand{\IJ}{_{12}}

\newcommand{\IM}{_{-1}}
\newcommand{\JM}{_{-2}}
\newcommand{\IMIM}{_{-1-1}}
\newcommand{\JMJM}{_{-2-2}}
\newcommand{\IMJM}{_{-1-2}}
\newcommand{\IMJ}{_{-12}}
\newcommand{\IJM}{_{1-2}}
\newcommand{\tp}{\tilde{\Psi}}
\newcommand{\T}{\theta}

\def\mref#1{(\ref{#1})}
\begin{document}



\title{
Integrable dynamics of Toda-type \\
on the square and triangular lattices}

\author{
P.M. Santini\thanks{Dipartimento di Fisica, Universit\`a di Roma
``La Sapienza'' and
Istituto Nazionale di Fisica Nucleare, Sezione di Roma,
Piazz.le Aldo Moro 2, I--00185 Roma, Italy,
e-mail: {\tt paolo.santini@roma1.infn.it}}~,
A. Doliwa\thanks{Uniwersytet Warmi\'{n}sko-Mazurski w Olsztynie,
Wydzia{\l} Matematyki i Informatyki,
ul.~\.{Z}o{\l}nierska 14 A, 10-561 Olsztyn, Poland
e-mail: {\tt doliwa@matman.uwm.edu.pl}},~and
M. Nieszporski\thanks{
Katedra Metod Matematycznych Fizyki,
Uniwersytet Warszawski
ul. Ho\.za 74, 00-682 Warszawa, Poland,
 e-mail: {\tt maciejun@fuw.edu.pl};
Department of Applied Mathematics, University of Leeds, Leeds LS2 9JT, UK
e-mail: maciejun@maths.leeds.ac.uk, tel: +44 113 343 5149 fax: +44 113 343 5090
}
}

\date\today

\maketitle

\begin{abstract}
In a recent paper we 
constructed an integrable generalization of the Toda law on the square lattice. In this 
paper we construct other examples of integrable dynamics of Toda-type 
on the square lattice, as well as on the triangular lattice, as    
nonlinear symmetries of the 
 discrete Laplace 
 equations on the square and
triangular lattices. We also construct 
the $\tau$ - function formulations and the Darboux-B\"acklund 
transformations of these novel dynamics.
  
\end{abstract}

\section{Introduction}
The Toda lattice \cite{Toda1,Toda2,Toda3}
\begin{equation} 
\label{Toda}
\frac{d^2 q_{m}}{dt^2}=\Delta_me^{\Delta_m q_{m-1}},
\end{equation}
where $\Delta_mf_m=f_{m+1}-f_m$ is the difference operator and $q_m(t)$ is a 
dynamical function on a one dimensional lattice, 
is one of the most famous integrable nonlinear lattice equations. 
It describes the dynamics of a one - dimensional physical lattice, the masses of which are 
subjected to an interaction potential of exponential type. 
The infinite, finite and periodic Toda lattice (\ref{Toda}), as well as 
its numerous extensions \cite{Hir,Levi,Date,Hirota2,Mikhailov,Bruschi,Rui,Sur,Yam}, 
have applications in various other physical and 
mathematical contexts \cite{Hiro,HiroS,Perk,Gib,Lu1,Lu2,Lukas}.

Motivated by these results we find it important to construct integrable generalizations of the Toda law (\ref{Toda})
to regular planar lattices; i.e., to the square, triangular and honeycomb lattices. 
 Since the spectral
problem associated with (\ref{Toda}) is an 
``integrable'' discretization of the one-dimensional Schr\"odinger operator \cite{Flaschka,Manakov}
(where by integrable we mean that the operator admits, as its continuous counterpart, a large set 
of continuous and discrete symmetries, like the Laplace and Darboux transformations 
(DTs)), such a project requires  the identification of proper integrable
discretizations of self-adjoint second order operators on the plane first.
A key progress in this direction 
was made in \cite{Nov}, where it was established that the self-adjoint scheme on the star of the triangular 
lattice admits Laplace transformations, and in
\cite{NSD,DNS}, where it was established that the self-adjoint 
schemes on the stars of the square, triangular and honeycomb lattices admit DTs, as  
their natural continuous counterparts. 
In addition, in \cite{DNS}, a novel discrete time dynamics on the triangular 
lattice was introduced, in connection with its Laplace transformation.
To construct integrable nonlinear dynamics associated with these self-adjoint operators,
gauge equivalent to the discrete Laplace equations on weighted graphs,
is the main goal of the paper.

It is  necessary to mention that these three planar schemes  (on the square, triangular and honeycomb lattices) are 
directly connected (see \cite{DGNS} and \cite{DNS}), via the sublattice approach \cite{DGNS},
to the so-called discrete Moutard \cite{DJM,NS} (or $B$-quadrilateral \cite{Doliwa})
lattice in $\ZZ^N$, and therefore they are all reductions of the 
multidimensional (planar) quadrilateral lattice \cite{DS1,DS2,DS3,DSManas}. 
We  also remark
that the above three linear schemes are distinguished examples of Laplace equations on graphs, 
obtainable from the discrete Moutard equations on 
bipartite planar quad-graphs \cite{Mer,K,BS1,BMS}.       

Using the self-adjoint scheme on the star of the square lattice as spectral operator, 
we have recently constructed in \cite{SND} the following example 
of integrable Toda-type dynamics on the square lattice
(together with its associated $\tau$ - function 
formulation, its Darboux and Darboux-B\"acklund transformations and some examples of explicit solutions):
\beq
\label{Toda_square_a}
\ba{l}
\xi_{m,n}\frac{d}{d t}\left(  \frac{1}{\xi_{m,n}}  \frac{dq_{m,n}}{d t}\right)=
\Delta_m\left( \xi_{m,n}\xi_{{m-1,n}}e^{\Delta_m q_{m-1,n}}\right)+
\Delta_n\left( \xi_{m,n}\xi_{{m,n-1}}e^{\Delta_n q_{m,n-1}}\right),  \\
\frac{\xi_{m,n}}{\xi_{m+1,n+1}}=e^{\Delta_m\Delta_nq_{m,n}}.
\ea
\eeq
where $q_{m,n}(t),\xi_{m,n}(t)$ are dynamical functions on the square lattice.

Motivated by the above results, in this paper we construct and study other examples of integrable dynamics 
of Toda-type on the square lattice, as well as on 
the triangular  lattice. In addition, we present their $\tau$ function formulations, 
in which the $\tau$ function of the BKP hierarchy \cite{Miwa} plays a central role, due to the 
already mentioned common origin of these schemes.
The integrability of the dynamics in question manifests here in the construction of the Lax pair and 
Darboux - B\"acklund transformations (DBTs).

We remark that, due to the  intimate connections between the self-adjoint  
schemes on the triangular and honeycomb lattices \cite{DNS},
it is possible, in principle, to construct integrable Toda-type dynamics on the honeycomb 
lattice from those on the triangular lattice.
This project will be developed elsewhere. Another interesting problem for future research is to 
establish connections between these Toda-like
systems and the corresponding Lotka-Volterra systems (see e.g. \cite{Nimmo,Gilson}), as well as 
the connection, via the sublattice approach, between these Toda-like
systems and the integrable dynamics on the
discrete Moutard lattice introduced in \cite{Kri}. 

The paper is organized as follows. In \S \ref{2} we construct an integrable dynamics 
of Toda - type on the 
square lattice, invariant  under $\pi/2$ - rotation, its 
$\tau$ - function formulation, and its two natural reductions transforming 
into each other under a $\pi/2$ - rotation. 
One of these two 
reductions coincides with the 2D Toda 
system (\ref{Toda_square_a}) introduced in \cite{SND}. 
In \S \ref{3} we construct an integrable dynamics of Toda 
type on the triangular lattice, invariant under a $\pi/3$ - rotation, its 
$\tau$ - function formulation and its
 natural reductions.
The DBTs for all the above systems 
are presented in \S \ref{4}.

\section{Dynamics on the square lattice}
\label{2}
In this section we construct examples of integrable dynamics of Toda - type 
on the square lattice. To simplify the form of the equations,from now on, we will be using the following notation:
$f$ instead of $f_{m,n}$, $f_{\pm 1}$ instead of $f_{m \pm 1,n}$, $f_{\pm 2}$ instead of $f_{m,n \pm 1}$, 
$f_{\pm 1\pm 2}$ instead of $f_{m \pm 1,n\pm 1}$,
$f_{\pm 1\pm1}$ instead of $f_{m \pm 2,n}$ and $f_{\pm 2\pm2}$ instead of $f_{m ,n \pm 2}$. Moreover we denote by $T_1$
and $T_2$ the basic translation operators acting on the lattice, i.e. 
$T_i f = f_i$, $i=1,2$.

\begin{figure}
\begin{center}
\mbox{ \epsfxsize=6cm \epsffile{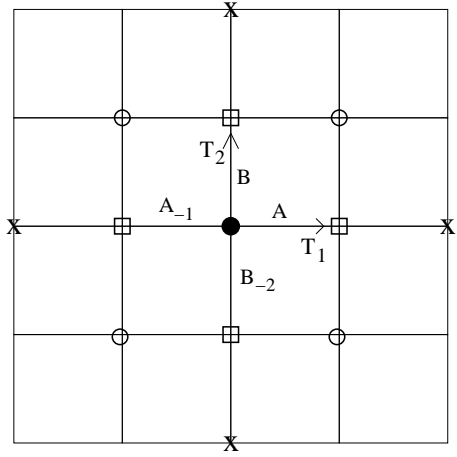}}
\end{center}
\caption{The square lattice and the points involved in the commutation.}
\label{fig-5s}
\end{figure}

The wanted dynamics are associated with the linear self-adjoint 5-point scheme
\beq
\label{5pt}
\ba{l} 
A\Psi_1+A_{-1}\Psi_{-1}+ B\Psi_{2}+B_{-2}\Psi_{-2}=F\Psi 
\ea 
\eeq
on the star of the square lattice, involving its black centre $\bullet$
and the four vertices of the star, denoted by the symbol $\Box$ in Fig \ref{fig-5s}.
In equation (\ref{5pt}) the eigenfunction $\Psi$ is defined at the vertices of the
graph, while the fields $A,B$ are defined on the non oriented edges 
of the lattice. Equation (\ref{5pt}), a natural discretization of the self-adjoint second order equation  
\beq
(a\psi_x)_x+(b\psi_y)_y=f\psi,
\eeq 
 admits, like its continuous counterpart, 
DTs \cite{NSD}.

We restrict our investigation to evolution equations for $\Psi$ involving only 
the $4$ vertices $\Box$ of the 5-point scheme:
\beq
\label{psi_evol_square}
\frac{d \Psi}{d t}=\alpha\Psi_1+\beta\Psi_{-1}+\gamma\Psi_2+\delta\Psi_{-2},
\eeq
where the fields $\alpha,\beta,\gamma,\delta$, defined on the oriented
edges of the lattice, 
will be specified in the following.  
A term proportional to $\Psi$ in (\ref{psi_evol_square}) can always be 
expressed, using (\ref{5pt}), in terms of the values of $\Psi$ at the $4$  
vertices $\Box$ of the star; therefore it is omitted.

We remark that, due to the $\pi/2$ - rotation symmetry of the square lattice, under 
which the two basic translations $T_1,T_2$ transform as follows: 
\beq
\label{transf_T_square}
T_1\to \tilde T_1=T_2,~T_2\to \tilde T_2=T^{-1}_1,
\eeq
the coefficients of the $5$-point scheme and of the evolution equation (\ref{psi_evol_square}) 
are subjected to the following transformations:
\beq\label{transf_coeff_square}
\ba{l}
A\to \tilde A=B,~~~B\to \tilde B=A_{-1},~~~F\to\tilde F=F,  \\
\alpha\to\tilde\alpha=\gamma,~~~\beta\to \tilde\beta=\delta,~~~
\gamma\to \tilde\gamma=\beta,~~~\delta\to \tilde\delta=\alpha.
\ea
\eeq
 
The compatibility between equations (\ref{5pt}) and (\ref{psi_evol_square}) 
leads to an equation involving the values of $\Psi$ at all the marked points 
$\circ,~\bullet,~\Box,~X$, in Fig.1. 
Using the scheme (\ref{5pt}) centered at the origin and at the points $\Box$, one 
expresses the values of $\Psi$ at the origin and at the points $X$ in terms of the $8$ 
independent values of $\Psi$ at the points $\Box$ and $\circ$. As a result of this 
procedure, the compatibility condition becomes a linear equation for the $8$ 
independent values of $\Psi$ at the points $\Box$ and $\circ$.     
Equating to zero their $8$ coefficients, one obtains  
a  determined system of $8$ nonlinear equations for the $8$ coefficients 
$A,B,C,F,\alpha,\beta,\gamma,\delta$. 

In the rest of this section we report the results of the analysis  of such system leading to Toda-type dynamics. 
\subsection{Rotationally invariant dynamics} 

Setting 
\beq\label{alpha_square}
\alpha=\frac{\xi}{2}A,~~~\beta=-\frac{\xi}{2}A_{-1},~~~\gamma=\frac{\eta}{2}B,
~~~\delta=-\frac{\eta}{2}B_{-2},
\eeq
where $\xi,\eta$ are lattice fields to be specified, the corresponding evolution for $\Psi$
\beq
\label{psi_evol_square_2}
\frac{d \Psi}{d t}=\frac{\xi}{2}\left(A\Psi_1-A_{-1}\Psi_{-1}\right)+
\frac{\eta}{2}\left(B\Psi_2-B_{-2}\Psi_{-2}\right)
\eeq
is compatible with the 5-point scheme (\ref{5pt}) iff (up to a trivial gauge transformation) 
the coefficients $A,B,F,\xi,\eta$ satisfy the following determined system of 5 nonlinear equations:  
\beq\label{dyn_square}
\ba{l}
\frac{dF}{dt}+\xi_1A^2-\xi_{-1}A^2_{-1}+\eta_2B^2-\eta_{-2}B^2_{-2}=0, \\
\;  \\
\frac{1}{A}\frac{d A}{dt}+\frac{1}{2}\Delta_1(\xi F)=0,~~~~~~~~~~~~~~~~~~~
\frac{1}{B}\frac{d B}{dt}+\frac{1}{2}\Delta_2(\eta F)=0, \\
\;  \\
AB(\xi+\eta)=A_2B_1(\xi+\eta)_{12},~~~~~~~~A_2B(\xi-\eta)_2=AB_1(\xi-\eta)_1.
\ea
\eeq

Equations (\ref{dyn_square}b),(\ref{dyn_square}c) suggest the introduction 
of the new fields $q,r$ defined by 
\beq\label{qr_def}
\frac{dq}{dt}=-\xi F,~~~\frac{dr}{dt}=-\eta F.
\eeq
With this choice:
\beq\label{AB}
A=ae^{\frac{1}{2}\Delta_1q},\;\;\;\;B=be^{\frac{1}{2}\Delta_2r},
\eeq
where $a,b$ are arbitrary constants. Choosing, w.l.g., $a,b=1$, 
the system (\ref{dyn_square}) takes the form of the following novel integrable 
generalization of the Toda law to the square lattice:
\beq\label{Toda_square}
\ba{l}
\xi\eta\frac{d}{d t}\left(\frac{1}{\xi}\frac{dq}{dt}\right)
=\eta\Delta_1\left(\xi\xi_{-1}e^{\Delta_1q_{-1}} \right)+
\xi\Delta_2\left(\eta\eta_{-2}e^{-\Delta_2r_{-2}} \right),                  \\
\xi \frac{dr}{dt}=\eta \frac{dq}{dt},                  \\
\frac{(\xi+\eta)}{(\xi+\eta)_{12}}=e^{\Delta_1\Delta_2\left(\frac{q+r}{2}\right)},~~~~~~~~
\frac{(\xi-\eta)_1}{(\xi-\eta)_{2}}=e^{\Delta_1\Delta_2\left(\frac{q-r}{2}\right)}.
\ea
\eeq

\vskip 10pt
\noindent
{\bf Remark 1}. In the natural 1-dimensional limit in which all the fields are invariant 
under the $T_2$ translation, equations (\ref{Toda_square}c,d) imply that $\xi$ and $\eta$ 
are constant, and equation (\ref{Toda_square}a) 
reduces to the 1-dimensional Toda lattice (\ref{Toda}).

\vskip 10pt
\noindent
{\bf Remark 2} Using (\ref{AB}) with $a=b=1$, the 5-point scheme (\ref{5pt}) takes the 
following form:
\beq\label{dSchrodinger}
\ba{l}
\frac{\Gamma}{\Gamma_1}\Psi_1+\frac{\Gamma_{-1}}{\Gamma}\Psi_{-1}+
\frac{\hat\Gamma}{\hat\Gamma_2}\Psi_2+\frac{\hat\Gamma_{-2}}{\hat\Gamma}\Psi_{-2}=F\Psi, \\
\Gamma=e^{-\frac{q}{2}},~~~~~\hat\Gamma=e^{-\frac{r}{2}}.
\ea
\eeq 
It is easy to verify that the spectral problem (\ref{dSchrodinger}) reduces, in the 
natural continuous limit, to the stationary Schr\"odinger equation in the plane: 
$\Psi_{xx}+\Psi_{yy}+u\Psi=0$. It is therefore a natural integrable discretization 
of the Schr\"odinger operator, more general than that introduced in \cite{NSD}.   

\vskip 10pt
\noindent
{\bf Remark 3} Using (\ref{transf_T_square}), (\ref{transf_coeff_square}), (\ref{alpha_square}) 
and (\ref{qr_def}), it is easy to verify that, under a $\pi/2$ - rotation, 
\beq\label{transf_coeff_square_2}
\xi\to\xi'=\eta,~~~~~\eta\to\eta'=-\xi,~~~~~ q\to q'=r,~~~~~r\to r'=-q;
\eeq 
from which it follows that the system (\ref{dyn_square}) (or (\ref{Toda_square})) is invariant 
under this transformation.
\subsection{Reductions not invariant under rotation}

The system (\ref{dyn_square}) (or (\ref{Toda_square})) admits two distinguished reductions 
for $\xi=\pm\eta$.
 
\vskip 10pt
\noindent
{\bf 1. The reduction $\xi=\eta$}. In this case, the Lax pair (\ref{5pt}),(\ref{psi_evol_square_2}) reduces 
to 
\beq
\label{Lax_square_a}
\ba{l}
A\Psi_1+A_{-1}\Psi_{-1}+ B\Psi_{2}+B_{-2}\Psi_{-2}=F\Psi,   \\
\frac{d \Psi}{d t}=\frac{\xi}{2}\left(A\Psi_1-A_{-1}\Psi_{-1}+B\Psi_2-B_{-2}\Psi_{-2}\right).
\ea
\eeq
and the nonlinear dynamics (\ref{dyn_square}) reduces to
\beq
\label{dyn_square_a}
\ba{l}
\frac{dF}{dt}+\xi_1A^2-\xi_{-1}A^2_{-1}+\xi_2B^2-\xi_{-2}B^2_{-2}=0, \\
\;  \\ 
\frac{1}{A}\frac{d A}{d t}+\frac{1}{2}\Delta_1(\xi F)=0,~~~~~~
\frac{1}{B}\frac{d B}{d t}+\frac{1}{2}\Delta_2(\xi F)=0, \\
\;  \\
AB\xi=A_2B_1\xi_{12};
\ea
\eeq
Integrating equations (\ref{dyn_square_a}b), (\ref{dyn_square_a}c) and using (\ref{qr_def}), 
which implies that $r=q~(\hat\Gamma=\Gamma)$, one recovers  
the Toda type system (\ref{Toda_square_a}), rewritten here, for completeness, in the new notation:
\beq
\label{Toda_square_aa}
\ba{l}
\xi\frac{d}{d t}\left(\frac{1}{\xi}\frac{dq}{dt}\right)=
\Delta_1\left( \xi\xi_{-1}e^{\Delta_1 q_{-1}}\right)+
\Delta_2\left( \xi\xi_{-2}e^{\Delta_2 q_{-2}}\right), \\
\frac{\xi}{\xi_{12}}=e^{\Delta_1\Delta_2q},
\ea
\eeq
and the associated 5-point scheme is the discrete Schr\"odinger equation
\beq\label{dSchrodinger_a}
\ba{l}
\frac{\Gamma}{\Gamma_1}\Psi_1+\frac{\Gamma_{-1}}{\Gamma}\Psi_{-1}+
\frac{\Gamma}{\Gamma_2}\Psi_2+\frac{\Gamma_{-2}}{\Gamma}\Psi_{-2}=F\Psi,  \\

A=\frac{\Gamma}{\Gamma_1},~~B=\frac{\Gamma}{\Gamma_2},~~\Gamma=e^{-\frac{q}{2}},~~F=-\frac{q}{\xi},
\ea
\eeq 
introduced in \cite{NSD}.

\vskip 10pt
\noindent
{\bf 2. The reduction $\xi=-\eta$}. In this case, the time evolution of $\Psi$ reads
\beq\label{psi_evol_square_b}
\frac{d \Psi}{d t}=\frac{\xi}{2}\left(A\Psi_1-A_{-1}\Psi_{-1}-B\Psi_2+B_{-2}\Psi_{-2}\right),
\eeq
and the nonlinear dynamics (\ref{dyn_square}) reduces to:
\beq\label{dyn_square_b}
\ba{l}
\frac{dF}{dt}+\xi_1A^2-\xi_{-1}A^2_{-1}-\xi_2B^2+\xi_{-2}B^2_{-2}=0, \\
\;  \\
\frac{1}{A}\frac{d A}{d t}+\frac{1}{2}\Delta_1(\xi F)=0,~~~~~~~
\frac{1}{B}\frac{d B}{d t}-\frac{1}{2}\Delta_2(\xi F)=0, \\
\;  \\
A_2B\xi_2=AB_1\xi_{1}.
\ea
\eeq
Equivalently, using (\ref{qr_def}) and noting that, in this case, $r=-q~(\hat\Gamma=1/\Gamma)$, 
one obtains the  Toda - type system:
\beq\label{Toda_square_b}
\ba{l}
\xi\frac{d}{d t}\left(\frac{1}{\xi}\frac{dq}{dt}\right)=\Delta_1\left(\xi\xi_{-1}e^{\Delta_1q_{-1}} \right)-
\Delta_2\left(\xi\xi_{-2}e^{-\Delta_2q_{-2}} \right),                  \\
\frac{\xi_1}{\xi_{2}}=e^{\Delta_1\Delta_2q},
\ea
\eeq
whose 5-point scheme is another variant of the discrete Schr\"odinger equation:
\beq\label{dSchrodinger_b}
\ba{l}
\frac{\Gamma}{\Gamma_1}\Psi_1+\frac{\Gamma_{-1}}{\Gamma}\Psi_{-1}+
\frac{\Gamma_2}{\Gamma}\Psi_2+\frac{\Gamma}{\Gamma_{-2}}\Psi_{-2}=F\Psi.
\ea
\eeq 

We end this section remarking that, due to the transformations (\ref{transf_coeff_square}), 
(\ref{transf_coeff_square_2}), the  reduced systems (\ref{Toda_square_aa}) and 
(\ref{Toda_square_b}) transform into each other under a $\pi/2$ - rotation.

\subsection{$\tau$ - function formulations}
Motivated by the sublattice approach \cite{DGNS} for the self-adjoint 5-point
scheme \eqref{5pt}, we introduce two potentials $\tau$ and $\hat\tau$ via
equations 
\beq A=\frac{\tau_1 \tau}{\hat\tau \hat\tau_{-2}}, \qquad
B=\frac{\tau_2 \tau}{\hat\tau \hat\tau_{-1}}.
\eeq
These allow to resolve the algebraic part (\ref{dyn_square}c) of the dynamic
equations (\ref{dyn_square}), with the fields $\xi$ and $\eta$ expressed
as follows
\beq \xi = \frac{\hat\tau_{-1}\hat\tau_{-2}+\hat\tau_{-1-2}\hat\tau}{2\tau^2}, 
\qquad
\eta =  \frac{\hat\tau_{-1}\hat\tau_{-2}-\hat\tau_{-1-2}\hat\tau}{2\tau^2} .
\eeq
Then the remaining equations (\ref{dyn_square}a) and (\ref{dyn_square}b) form a
system of three equations for three fields $\tau$, $\hat\tau$ and $F$
\bea
4\frac{d}{d t}\left( \log \frac{\tau_1 \tau}{\hat\tau \hat\tau_{-2}} \right) +
\Delta_1 \left[ F \left( \frac{\hat\tau_{-1}\hat\tau_{-2}}{\tau^2} +
\frac{\hat\tau_{-1-2}\hat\tau}{\tau^2} \right) \right] = 0, \nonumber \\
4 \frac{d}{d t}\left( \log \frac{\tau_2 \tau}{\hat\tau \hat\tau_{-1}} \right) +
\Delta_2 \left[ F \left( \frac{\hat\tau_{-1}\hat\tau_{-2}}{\tau^2} -
\frac{\hat\tau_{-1-2}\hat\tau}{\tau^2} \right) \right] = 0,\nonumber \\
\frac{2}{\tau^2} \frac{dF}{dt} + \frac{1}{\hat\tau \hat\tau_{-2}}
\left( \frac{\hat\tau_{1-2}}{\hat\tau_{-2}} + \frac{\hat\tau_{1}}{\hat\tau} 
\right)
- \frac{1}{\hat\tau_{-1} \hat\tau_{-1-2}}
\left( \frac{\hat\tau_{-1-1}}{\hat\tau_{-1}} + \frac{\hat\tau_{-1-1-2}}
{\hat\tau_{-1-2}}
\right) + \nonumber \\ 
\frac{1}{\hat\tau \hat\tau_{-1}}
\left( \frac{\hat\tau_{-12}}{\hat\tau_{-1}} - \frac{\hat\tau_{2}}{\hat\tau} 
\right) - \frac{1}{\hat\tau_{-2} \hat\tau_{-1-2}}
\left( \frac{\hat\tau_{-2-2}}{\hat\tau_{-2}} - \frac{\hat\tau_{-1-1-2}}
{\hat\tau_{-1-2}} \right) =0.    \nonumber  
\eea

Introduction of the
fields $q$ and $r$ (or $\Gamma$ and $\hat\Gamma$), which allowed
to simplify equations (\ref{dyn_square}b), suggests the introduction of yet other
potentials $h$ and $\hat{h}$ such that
\beq
\tau^2 = \frac{\hat{h}_1 \hat{h}_2}{h h_{12}} ,\qquad
\hat\tau = \left( \frac{\hat{h}}{h} \right)_{12}.
\eeq
It follows that
\beq
\Gamma^2 = \frac{h_{12}\hat{h}_2}{h \hat{h}_1}, \qquad 
\hat\Gamma^2 = \frac{h_{12}\hat{h}_1}{h \hat{h}_2},
\eeq 
and that equations (\ref{dyn_square}c) are identically satisfied,
with the fields $\xi$ and $\eta$ given as follows
\beq \xi = \frac{1}{2}\left( \frac{h h_{12}}{h_1 h_2} +
\frac{\hat{h}\hat{h}_{12}}{\hat{h}_{1}\hat{h}_{2}} \right), \qquad
\eta = \frac{1}{2}\left( \frac{h h_{12}}{h_1 h_2} -
\frac{\hat{h}\hat{h}_{12}}{\hat{h}_{1}\hat{h}_{2}} \right).
\eeq 
Moreover, equations (\ref{dyn_square}b) reduce to two equivalent expressions
for $F$
\beq \label{Toda7-2f-1}
 \frac{h_1 h_2}{h h_{12}}\frac{d}{d t}\left( \log \frac{h_{12}}{h} \right) =
\frac{\hat{h}_{1}\hat{h}_{2}}{\hat{h}\hat{h}_{12}} 
\frac{d}{d t}\left( \log \frac{\hat{h}_{2}}{\hat{h}_{1}} \right) = \frac{F}{2},
\eeq 
while equation (\ref{dyn_square}c) reads 
\begin{multline} \label{Toda7-2f-2}
4\frac{d}{d t}\left(\frac{h_1 h_2}{h h_{12}}\frac{d}{d t}\left( \log \frac{h_{12}}{h} \right) \right)
+  \left(
\frac{h \hat{h}_1}{h_1 \hat{h}} +\frac{h_2 \hat{h}_{12}}{h_{12} \hat{h}_2} 
\right)_1 \frac{h_1 \hat{h}_2}{h \hat{h}_{12}} -
\left(
\frac{h_1 \hat{h}}{h \hat{h}_1} +\frac{h_{12} \hat{h}_{2}}{h_{2} \hat{h}_{12}} 
\right)_{-1} \frac{h_2 \hat{h}_1}{h_{12} \hat{h}} + \\
\left(
\frac{h \hat{h}_2}{h_2 \hat{h}} - \frac{h_1 \hat{h}_{12}}{h_{12} \hat{h}_1} 
\right)_2 \frac{h_2 \hat{h}_1}{h \hat{h}_{12}} -
\left(
\frac{h_{12} \hat{h}_{1}}{h_{1} \hat{h}_{12}} -
\frac{h_2 \hat{h}}{h \hat{h}_2}  
\right)_{-2} \frac{h_1 \hat{h}_2}{h_{12} \hat{h}} =0.
\end{multline}
Therefore the introduction of the potentials $h$ and $\hat{h}$ allows one to
rewrite the Toda-like system (\ref{dyn_square}) as a coupled nonlinear system of
two equations  (the first equation of \eqref{Toda7-2f-1} and equation
\eqref{Toda7-2f-2}).

\section{Dynamics on the triangular lattice}
\label{3}
In this section we construct some examples of integrable dynamics of Toda type on the regular
triangular lattice.  We recall  that, on the triangular lattice,
the three main translations $T_1,T_2,T_3$ in the directions $1$, $2$ and $3$
are not independent, being connected by the relation 
\beq\label{T_relation}
T_1T_3=T_2,
\eeq
and hence $f_3=f_{-12}$, $f_{-3}=f_{1-2}$.

\begin{figure}
\begin{center}
\mbox{ \epsfxsize=10cm \epsffile{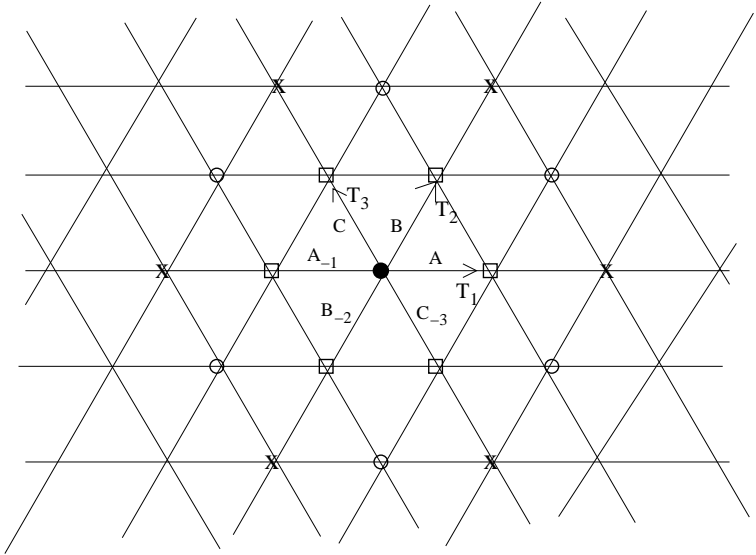}}
\end{center}
\caption{The triangular lattice and the points involved in the commutation}
\label{fig-7s}
\end{figure}

The integrable dynamics of Toda type are associated with the linear and self-adjoint 7-point scheme
\beq\label{7pt}
A\Psi_1+A_{-1}\Psi_{-1}+B\Psi_2+B_{-2}\Psi_{-2}+C\Psi_3+C_{-3}\Psi_{-3}=F\Psi, 
\eeq
on the star of the triangular lattice, involving the black centre $\bullet$ 
and the $6$ vertices denoted by the symbol $\Box$ in Fig. \ref{fig-7s}. 
In equation (\ref{7pt}) the eigenfunction $\Psi$ is defined at the vertices of the
graph and the fields $A,B,C$ are defined on the non oriented edges 
of the lattice. Equation (\ref{7pt}), a natural discretization of the most general self-adjoint 
second order equation on the plane: 
\beq
(a\psi_x)_x+(b\psi_y)_y+(c\psi_x)_y+(c\psi_y)_x=f\psi, 
\eeq 
admits, like its continuous counterpart, DTs \cite{NSD}.

As in the previous section, we restrict our investigation to evolution equations for $\Psi$ 
involving only the $6$ points $\Box$ of the 7-point scheme:
\beq\label{psi_evol_triangle}
\frac{d \Psi}{d t}=\alpha\Psi_1+\beta\Psi_{-1}+\gamma\Psi_2+\delta\Psi_{-2}+\epsilon\Psi_3+\nu\Psi_{-3} .
\eeq

We remark that the regular triangular lattice possesses a $\pi/3$ - rotation symmetry, under which 
the 3 basic translations transform as follows:
\beq\label{{transf_T_triangle}}
T_1\to \tilde T_1=T_2,~~~T_2\to \tilde T_2=T_3,~~~T_3\to \tilde T_3=T^{-1}_1.
\eeq
Correspondingly, the coefficients of the $7$ - point scheme (\ref{7pt}) and of the evolution 
equation (\ref{psi_evol_triangle}) transform as follows:
\beq
\label{transf_coeff_triangle}
\ba{l}
A\to B,~B\to C,~C\to A_{-1}, \\
\alpha\to\tilde\alpha=\gamma,~~~\beta\to \tilde\beta=\delta,~~~
\gamma\to \tilde\gamma=\epsilon,~~~\delta\to \tilde\delta=\nu,  \\
\epsilon\to \tilde\epsilon=\beta,~~~\nu\to \tilde\nu=\alpha.
\ea
\eeq

We proceed adopting the same strategy as in the previous section. The compatibility between 
equations (\ref{7pt}) and (\ref{psi_evol_triangle}) 
leads to an equation involving the values of $\Psi$ at all the $19$ marked points in Fig.2. 
Using the scheme (\ref{7pt}) centered at the origin and at the points $\Box$, one 
expresses the values of $\Psi$ at the origin and at the points $X$ in terms of the $12$ 
values of $\Psi$ at the points $\Box$ and $\circ$. As a result of this 
procedure, the compatibility condition becomes a linear equation for the $12$ 
independent values of $\Psi$ at the points $\Box$ and $\circ$.    
Equating to zero their $12$ coefficients, one obtains   
an  overdetermined system of $12$ nonlinear equations for the $10$ coefficients 
$A,B,C,F,\alpha,\beta,\gamma,\delta,\epsilon,\nu$. It turns out that, due to the relation (\ref{T_relation}) 
among the three main shifts, such overdeterminacy is resolved, and one construct integrable nontrivial 
dynamics.

In the rest of this section we report the results of such analysis,
leading to the Toda-type dynamics on the triangular lattice.
\subsection{Rotationally invariant dynamics} 

Setting 
\beq\label{alpha_triangle}
\alpha=\frac{\xi}{2}A,~\beta=-\frac{\xi}{2}A_{-1},~\gamma=\frac{\eta}{2}B,
~\delta=-\frac{\eta}{2}B_{-2},~\epsilon=\frac{\zeta}{2}C,~\nu=-\frac{\zeta}{2}C_{-3},
\eeq
where $\xi,\eta,\zeta$ are lattice fields to be specified, the corresponding 
evolution for $\Psi$
\beq\label{psi_evol_triangle_2}
\frac{d \Psi}{d t}=\frac{\xi}{2}\left(A\Psi_1-A_{-1}\Psi_{-1}\right)+
\frac{\eta}{2}\left(B\Psi_2-B_{-2}\Psi_{-2}\right)+
\frac{\zeta}{2}\left(C\Psi_3-C_{-3}\Psi_{-3}\right)
\eeq
is compatible with the 7-point scheme (\ref{7pt}) iff the coefficients 
$A,B,C,F,\xi,\eta,\zeta$ satisfy 
the following determined system of 7 nonlinear equations:  
\beq\label{dyn_triangle}
\ba{l}
\frac{dF}{dt}+\frac{1}{\xi}\Delta_1\left(\xi\xi_{-1}A^2_{-1}\right)+ 
\frac{1}{\eta}\Delta_2\left(\eta\eta_{-2}B^2_{-2}\right)+
\frac{1}{\zeta}\Delta_3\left(\zeta\zeta_{-3}C^2_{-3}\right)=0, \\
\;  \\
\frac{1}{A}\frac{d A}{d t}+\frac{1}{2}\Delta_1(\xi F)-
\frac{1}{2}\frac{BC_1}{A}(\eta+\zeta)_2+\frac{1}{2}\frac{B_{-3}C_{-3}}{A}(\eta+\zeta)_{-3}=0, \\
\;  \\
\frac{1}{B}\frac{d B}{d t}+\frac{1}{2}\Delta_2(\eta F)+
\frac{1}{2}\frac{A_3C}{B}(\xi-\zeta)_3-\frac{1}{2}\frac{AC_{1}}{B}(\xi-\zeta)_{1}=0, \\
\;  \\
\frac{1}{C}\frac{d C}{d t}+\frac{1}{2}\Delta_3(\zeta F)+
\frac{1}{2}\frac{B_{-1}A_{-1}}{C}(\xi+\eta)_{-1}-\frac{1}{2}\frac{A_3B}{C}(\xi+\eta)_{2}=0, \\
\;  \\
AB_1(\xi-\eta)_1=A_2B(\xi-\eta)_{2},~~\\
\;  \\
AC(\xi+\zeta)=A_3C_1(\xi+\zeta)_2,~~ \\
\;  \\
B_{-3}C_1(\eta-\zeta)_1=BC_{-3}(\eta-\zeta).
\ea
\eeq
We remark that, due to equations (\ref{dyn_triangle}e)-(\ref{dyn_triangle}g), 
equations  (\ref{dyn_triangle}b)-(\ref{dyn_triangle}d) can be rewritten in 
the following 
conservation-like form:
\beq\label{dyn_triangle_cons}
\ba{l}
\frac{d}{d t}(\ln A^2)+\Delta_1(\xi F)-
\Delta_2\left(\frac{B_{-2}C_{-3}}{A_{-2}}(\xi-\eta) \right)-
\Delta_3\left(\frac{B_{-3}C_{-3}}{A}(\xi+\zeta)_{-3} \right)=0, \\
\;  \\
\frac{d}{d t}(\ln B^2)+\Delta_2(\eta F)+
\Delta_3\left(\frac{AC_{-3}}{B_{-3}}(\eta-\zeta) \right)-
\Delta_1\left(\frac{A_{-1}C}{B_{-1}}(\xi-\eta) \right)=0, \\
\;  \\
\frac{d}{d t}(\ln C^2)+\Delta_3(\zeta F)-
\Delta_1\left(\frac{A_{-1}B_{-1}}{C}(\xi+\zeta)_{-1} \right)-
\Delta_2\left(\frac{A_{-1}B_{-2}}{C_{-2}}(\eta-\zeta) \right)=0. 
\ea
\eeq

\vskip 10pt
\noindent
{\bf Remark 4} Under the transformation (\ref{transf_coeff_triangle}), 
the coefficients $\xi,\eta,\zeta$ transform as follows:
\beq\label{transf_coeff_triangle_2}
\xi\to\tilde\xi=\eta,~~~\eta\to\tilde\eta=\zeta,~~~\zeta\to\tilde\zeta=-\xi,
\eeq
and, as it is easy to verify, the nonlinear system (\ref{dyn_triangle}) is invariant under 
a $\pi/3$ - rotation.

\subsection{Reductions not invariant under rotation}

\begin{figure}[h!]
\begin{center}
\mbox{ \epsfxsize=10cm \epsffile{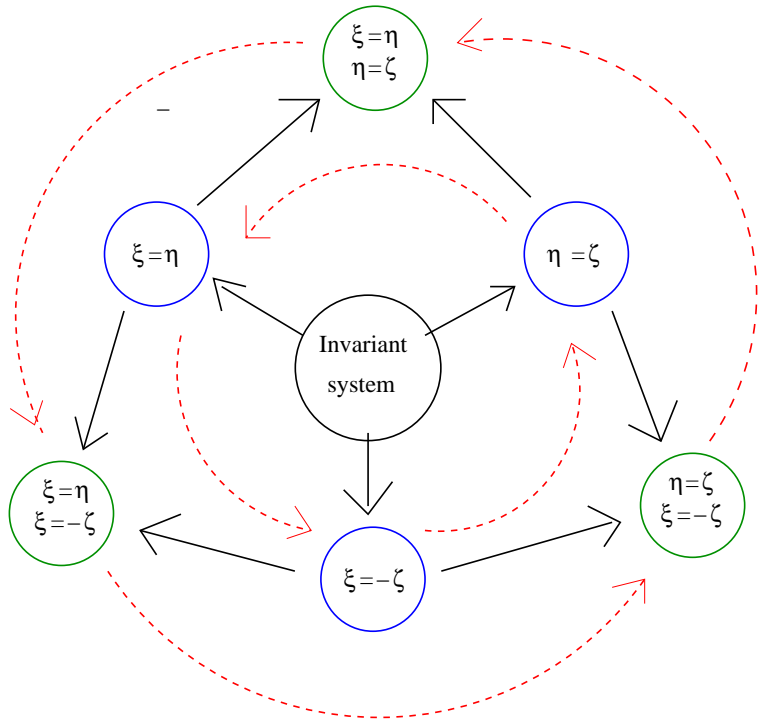}}
\end{center}
\caption{The bold arrows describe the 6 reductions of the rotationally invariant system.  
The dashed arrows describe how such reductions transform one into the other, under a $\pi/3$ 
rotation.} 
\label{fig-r}
\end{figure}

The nonlinear system  (\ref{dyn_triangle}) admits the   reductions $[\xi=\eta]$, 
$[\eta=\zeta]$ and $[\zeta=-\xi]$, and the following combinations of them: $[\xi=\eta,~\eta=\zeta]$, 
$[\xi=\eta,~\zeta=-\xi]$, and $[\eta=\zeta,~\zeta=-\xi]$. 
They give rise to six  integrable dynamics on the triangular lattice. 
It follows that these  dynamics are not rotationally invariant, but they transform one into the 
other in the way summarized in Fig. \ref{fig-r}.
\vskip 10pt
We write down explicitely the two reductions $[\xi=\eta]$ and $[\xi=\eta,~\eta=\zeta]$, since all 
the others can be generated from them through rotations.  
 
\vskip 10pt
\noindent
{\bf The reduction $\xi=\eta$}. In this case, the evolution of $\Psi$ reads:
\beq\label{psi_evol_triangle_red_1}
\frac{d \Psi}{d t}=\frac{\xi}{2}\left(A\Psi_1-A_{-1}\Psi_{-1}+B\Psi_2-B_{-2}\Psi_{-2}\right)+
\frac{\zeta}{2}\left(C\Psi_3-C_{-3}\Psi_{-3}\right),
\eeq
and the nonlinear system (\ref{dyn_triangle}) reduces to the six equations
\beq
\label{dyn_triangle_red_1}
\ba{l}
\frac{d F}{d t}+\frac{1}{\xi}\left( \Delta_1\left(\xi\xi_{-1}A^2_{-1}\right)+
\Delta_2\left(\xi\xi_{-2}B^2_{-2}\right)\right)+
\frac{1}{\zeta}\Delta_3\left(\zeta\zeta_{-3}C^2_{-3}\right)=0, \\
\;  \\
\frac{d (\ln A^2)}{d t}+\Delta_1(\xi F)-
\Delta_3\left(\frac{B_{-3}C_{-3}}{A}(\xi+\zeta)_{-3} \right)=0, \\
\;  \\
\frac{d(\ln B^2)}{d t}+\Delta_2(\xi F)+
\Delta_3\left(\frac{AC_{-3}}{B_{-3}}(\xi-\zeta) \right)=0, \\
\;  \\
\frac{d(\ln C^2)}{d t}+\Delta_3(\zeta F)-
\Delta_1\left(\frac{A_{-1}B_{-1}}{C}(\xi+\zeta)_{-1} \right)-
\Delta_2\left(\frac{A_{-1}B_{-2}}{C_{-2}}(\xi-\zeta) \right)=0. \\
\;  \\
AC(\xi+\zeta)=A_3C_1(\xi+\zeta)_2,~~ \\
\;  \\
B_{-3}C_1(\xi-\zeta)_1=BC_{-3}(\xi-\zeta).
\ea
\eeq
   
\vskip 10pt
\noindent
{\bf The reduction ($\xi=\eta,~\eta=\zeta$)}. In this case, the evolution of $\Psi$ reads:
\beq
\label{psi_evol_triangle_red_2}
\frac{d\Psi}{d t}=\frac{\xi}{2}\left(A\Psi_1-A_{-1}\Psi_{-1}+B\Psi_2-B_{-2}\Psi_{-2}+
C\Psi_3-C_{-3}\Psi_{-3}\right),
\eeq
and the nonlinear system (\ref{dyn_triangle}) reduces to the five equations
\beq
\label{dyn_triangle_red_2}
\ba{l}
\xi \frac{d F}{d t}+\Delta_1\left(\xi\xi_{-1}A^2_{-1}\right)+
\Delta_2\left(\xi\xi_{-2}B^2_{-2}\right)+
\Delta_3\left(\zeta\zeta_{-3}C^2_{-3}\right)=0, \\
\;  \\
\frac{d (\ln A^2)}{d t}+\Delta_1(\xi F)-
2\Delta_3\left(\frac{B_{-3}C_{-3}}{A}\xi_{-3} \right)=0, \\
\;  \\
\frac{d (\ln B^2)}{d t}+\Delta_2(\xi F)=0, \\
\;  \\
\frac{d (\ln C^2)}{d t}+\Delta_3(\xi F)-
2\Delta_1\left(\frac{A_{-1}B_{-1}}{C}\xi_{-1} \right)=0, \\
\;  \\
AC\xi=A_3C_1\xi_2.
\ea
\eeq

In analogy with the previous examples on the square lattice, equations 
(\ref{dyn_triangle_red_2}b) - (\ref{dyn_triangle_red_2}d) suggest the introduction of the new fields 
$q,\rho,\sigma$ defined by 
\beq\label{qrs-def}
\frac{d q}{d t}=-\xi F,\;\;\;\;
\frac{d \rho}{d t}=-2\frac{BC}{A_3}\xi,\;\;\;\;
\frac{d \sigma}{d t}=-2\frac{AB}{C_1}\xi.
\eeq
With this choice:
\beq
A=ae^{\frac{1}{2}(\Delta_1q-\Delta_3\rho_{-3})},\;\;\;\;B=be^{\frac{1}{2}\Delta_2q},\;\;\;\;
C=ce^{\frac{1}{2}(\Delta_3q-\Delta_1\sigma_{-1})},
\eeq
where $a,b,c$ are arbitrary constants. Choosing, w.l.g., $a,b,c=1$, 
the system (\ref{dyn_triangle_red_2}) can be rewritten in the following Toda-like form:
\beq\label{Toda_triangle_red_2}
\ba{l}
\xi\frac{d}{d t}\left(\frac{1}{\xi}\frac{d q}{d t}\right)+\Delta_1\left(\xi\xi_{-1}e^{\Delta_1q_{-1}+\rho_{-2}-r_{-1}} \right)+
\Delta_2\left(\xi\xi_{-2}e^{\Delta_2q_{-2}} \right)+                                  \\
\;  \\
\Delta_3\left(\xi\xi_{-3}e^{\Delta_3q_{-3}+\rho_{-2}-\sigma_{-3}} \right)=0,                  \\
\;  \\
\frac{d\rho}{d t}=-2\xi e^{\Delta_3(q+\frac{\rho}{2})-\frac{1}{2}\Delta_1\sigma_{-1} },                    \\
\;  \\
\frac{d\sigma}{d t} =-2\xi e^{\Delta_1(q+\frac{\sigma}{2})-\frac{1}{2}\Delta_3\rho_{-3} },                    \\
\;  \\
\xi_2=\xi e^{q_3-q_2+q_1-q+\frac{1}{2}(\Delta^2_1\sigma_{-1}+\Delta^2_3\rho_{-3})}.
\ea
\eeq

\subsection{Reductions to dynamics on the $\ZZ^2$ graph}

We remark that the reduction $\xi=\eta,~\eta=\zeta$ is compatible with the 
condition $B=0$, for which  all the connections in the direction $2$    
are broken, and the triangular lattice reduces to the rhombic lattice in Fig.4a. 
Then the direction $2$ should be renamed $13$ and the system (\ref{dyn_triangle_red_2}) 
becomes the integrable system (\ref{dyn_square_a}) on the rhombic lattice of Fig.4a 
(on the $\ZZ^2$ graph).

\begin{figure}
\begin{center}
\mbox{ \epsfxsize=14cm \epsffile{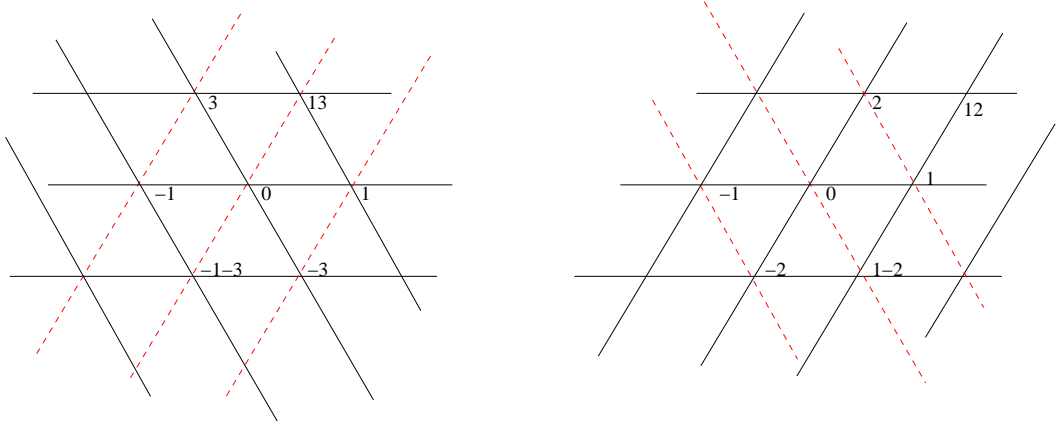}}
\caption{a) The rhombic lattice for $B=0$.\;\; b) The rhombic lattice for $C=0$.}
\label{fig-rho}
\end{center}
\end{figure}

Analogously, it would be possible to show, f.e., that the reduction $\xi=-\eta=-\zeta$ 
is compatible with the 
condition $C=0$, for which  all the connections in the direction $3$    
are broken, and the triangular lattice reduces to the rhombic lattice in Fig.\ref{fig-rho}b. The system 
obtained in this case is the integrable system (\ref{dyn_square_b}) on such a rhombic lattice. 

\subsection{$\tau$ - function formulations}
Motivated by the sublattice approach \cite{DNS} for the self-adjoint 7-point
scheme \eqref{7pt}, we introduce three potentials $\tau$, $\hat\tau$ and
$\check\tau$ such that
\beq
A = \frac{\tau \tau_1}{\hat\tau \check\tau_2}, \qquad
B = \frac{\tau \tau_2}{\hat\tau_3 \check\tau_2}, \qquad
C = \frac{\tau \tau_3}{\hat\tau_3 \check\tau_3}.
\eeq 
The algebraic part (\ref{dyn_triangle}e)-(\ref{dyn_triangle}g) of the
nonlinear system \eqref{dyn_triangle} is then resolved by the parametrization
\bea
\xi =  \frac{1}{2\tau^2}(  \check\tau_3 \hat\tau
+ \check\tau_2 \hat\tau_{-1} + \check\tau \hat\tau_3 ), \nonumber \\
\eta =  \frac{1}{2\tau^2}(  \check\tau_3 \hat\tau
- \check\tau_2 \hat\tau_{-1} + \check\tau \hat\tau_3 ), \\
\zeta =  \frac{1}{2\tau^2}(\check\tau_3 \hat\tau -
 \check\tau_2 \hat\tau_{-1} - \check\tau \hat\tau_3 ). \nonumber
\eea
The remaining part of the system gives a system of four equations for the four
fields $F$, $\tau$, $\check\tau$ and $\hat\tau$.

\section{Darboux - B\"acklund transformations}
In this section we present the DBTs for the above Toda-type systems.
\label{4}
\subsection{DBTs for the Toda-type system on the square lattice}

The Lax pair \eqref{5pt}, \eqref{psi_evol_square_2}
is covariant under the 
 Darboux transformation 
 \beq (\Psi, A, \dots, \eta) \mapsto (\tilde\Psi, \tilde{A}, \dots, \tilde\eta)
 \eeq
 given by the linear system
\begin{equation}
\begin{array}{l}
\Dj \left( K  \tp \right)= -A \IM \T \T \IM \Dim \left( \frac{\Psi}{\T}
\right),\\
\Di \left( K  \tp \right)= B \JM \T \T \JM \Djm \left( \frac{\Psi}{\T}
\right),\\
\frac{d}{d t}\left( K  \tp \right) = 
\T \IM \T \JM A \IM B \JM \frac{(\xi+\eta)}{2}
\left[ \left(\frac{\Psi}{\T}\right)\JM-\left(\frac{\Psi}{\T}\right)\IM
\right]+\\
\T  \T \IMJM A \IM B \IMJM \frac{(\xi \IM-\eta \IM)}{2}
\left[ \left(\frac{\Psi}{\T}\right)\IMJM -\frac{\Psi}{\T} \right],
\end{array}
\end{equation} 
where $\T$ is a particular solution of \mref{5pt},  
the gauge function $K$ must obey 
\begin{equation}
\begin{array}{l}
\frac{1}{K}\frac{d K}{d t}+ 
(\T \IM A \IM +\T \JM B \JM)\frac{\xi + \eta}{4 \T}
-
(\T  \JM A \IMJM+\T \IM B\IMJM)\frac{(\xi + \eta )\IMJM}{4 \T\IMJM}
+\\
(\T \IMJM A \IMJM+\T B \JM )\frac{(\xi - \eta )\JM}{4 \T\JM}
-(\T A \IM+\T \IMJM B \IMJM )\frac{(\xi - \eta )\IM}{4 \T\IM}
=0,
\end{array}
\end{equation}
and the transformation of the other fields is given by 
\begin{equation}
\begin{array}{l}
\label{DBt5}
\tilde{A}=\frac{K K \I}{B\JM \T \T \JM},\\
\tilde{B}=\frac{K K \J}{A\IM \T \T \IM},\\
\tilde{F}=K^2 (\frac{1}{A\IM\T\IM \T}+
\frac{1}{A\IMJM \T\IMJM \T \JM}+\frac{1}{B \JM \T \JM \T}+
\frac{1}{B \IMJM\T\IMJM \T \IM}),\\
\tilde{\xi}-\tilde{\eta}=\frac{\T \IMJM \T}{K^2 } A\IM B\IMJM (\xi
-\eta )_{-1},\\
\tilde{\xi}+\tilde{\eta}=\frac{\T \IM \T \JM}{K^2 }A \IM B\JM 
(\xi+\eta).
\end{array}
\end{equation}
So the formulae \mref{DBt5}
are
the B\"acklund transformations (BTs) for the Toda-type system 
(\ref{dyn_square}) on the square lattice, i.e. 
$\tilde{A}, \tilde{B}, \tilde{F}, \tilde{\xi}, \tilde{\eta}$ is a new solution of (\ref{dyn_square}).
On the level of the $\tau$-functions the transformation is given as follows
\beq \tau \mapsto K \hat\tau_{-1-2}, \qquad
\hat\tau \mapsto \theta\tau.
\eeq

We remark that, for $\xi =\pm \eta$, the above transformations become the DBTs for the reduced 
systems (\ref{dyn_square_a}) and (\ref{dyn_square_b}).

The spatial parts of the above DBTs were already written in \cite{NSD}; the temporal parts, describing the 
time dependence of the transformed solution $\tilde\Psi$, and the transformation 
law for the coefficient $\xi,\eta,\zeta$, are new ingredients of this paper.

\subsection{DBTs for the Toda type system on the triangular lattice}

First, for aesthetical reasons, we introduce function
\[S:=C \JM. \]
The Lax pair \eqref{7pt}, \eqref{psi_evol_triangle_2}
is covariant under the 
 Darboux transformation 
 \beq (\Psi, A, \dots, \zeta) \mapsto 
 (\tilde\Psi, \tilde{A}, \dots, \tilde\zeta)
 \eeq
 given by the linear system
\begin{equation}
\label{DBT7}
\begin{array}{l}
\Di (K  \tilde{\Psi}) =-B \JM \T \JM \Psi -S \T \JM \Psi \IM+ (B\JM \T + S \T \IM)\Psi \JM,\\
\Dj (K  \tilde{\Psi}) =A\IM \T \IM \Psi-(A\IM \T + S \T \JM) \Psi \IM+ S \T \IM
\Psi \JM, \\
\frac{d}{d t}(K  \tilde{\Psi})= 
-\frac{1}{2} \{\T\IMJM A\IM B\IMJM (\xi -\eta ) \IM \Psi + \\
\T \JM A \IMIM S (\xi + \zeta)\IM \Psi \IMIM
-\T \IM B \JMJM S (\eta - \zeta )\JM \Psi \JMJM 
+\\
~[\frac{P \JM B \IMJM}{B\JMJM} (\eta - \zeta ) \IMJM +
\T\IMJM S A\IMJM (\xi -\eta )\JM +
\T \JM A\IMJM B\IMJM (\xi  + \zeta ) \IMJM  ] \Psi \IM
-\\
~[\frac{P\IM A\IMJM}{A\IMIM} (\xi+ \zeta)\IMJM -\T\IMJM S B\IMJM(\xi -\eta )\IM +
\T \IM {A\IMJM B\IMJM}(\eta - \zeta )\IMJM] \Psi \JM 
-\\
~[\frac{P B\IMJM}{B\JM}(\xi -\eta)\IM-\T \JM S B \IMJM(\xi  + \zeta )\IM + 
\T \IM S A \IMJM (\eta - \zeta )\JM ] \Psi \IMJM \},
\end{array}
\end{equation}
where $\T$ is a particular solution of the Lax pair \eqref{7pt}, 
\eqref{psi_evol_triangle_2}, $P$ is given by
\[P:=\T A\IM B\JM +\T \IM A\IM S +\T \JM B\JM S, \]
and $K$ is given by the quadrature
\begin{equation}
\label{K}
\begin{array}{l}
\frac{1}{K}\frac{d K}{dt}=\frac{-1}{4}\{
P \T \IMJM \frac{B\IMJM}{B\JM}(\xi -\eta)_{-1}
[\frac{A \IM}{\T \JM P}-\frac{A \IMIM}{\T \IMJM P\IM}-\frac{B \JM}{\T \IM P}+\frac{B \JMJM}{\T \IMJM P\JM}
-\frac{S \IM}{\T \IM P\IM}+\frac{S \JM}{\T \JM P\JM}]+\\
P \IM \T \JM \frac{A\IMJM}{A\IMIM}(\xi  + \zeta )_{-1-2}
[\frac{A \IM}{\T \JM P}-\frac{A \IMIM}{\T \IMJM P\IM}+
\frac{B \JM}{\T \IM P}-\frac{B \JMJM}{\T \IMJM P\JM}
+\frac{S \IM}{\T \IM P\IM}-\frac{S \JM}{\T \JM P\JM}]
+\\
P \JM \T \IM \frac{B\IMJM}{B\JMJM}(\eta - \zeta )_{-1-2} 
[\frac{A \IM}{\T \JM P}-\frac{A \IMIM}{\T \IMJM P\IM}+\frac{B \JM}{\T \IM P}-\frac{B \JMJM}{\T \IMJM P\JM}
-\frac{S \IM}{\T \IM P\IM}+\frac{S \JM}{\T \JM P\JM}]\}.
\end{array}
\end{equation}

The new eigenfunction $\tilde{\Psi}$ is a solution of the
Lax pair \eqref{7pt}, 
\eqref{psi_evol_triangle_2} with the new coefficients
\begin{equation}
\begin{array}{l}
\label{DBt7}
\tilde{A}=\frac{K K\I}{\T \JM P} A\IM ,\\
\tilde{B}=\frac{K K\J}{\T \IM P} B\JM ,\\
\tilde{S}=\frac{K \IM K\JM}{\T \IMJM P\IMJM} S\IMJM, \\
\tilde{F}= K^2
\left(
\frac{A\IM}{\T \JM P}  +
\frac{A\IMIM}{\T \IMJM P\IM}+
\frac{B\JM}{\T \IM P}+
\frac{B\JMJM}{\T \IMJM P\JM}+
\frac{S\IM}{\T \IM P\IM}+
\frac{S\JM}{\T \JM P\JM}
\right)\\
\tilde{\xi}-\tilde{\eta}=\frac{P B\IMJM \T \IMJM}{K^2 B \JM} (\xi-\eta) \IM ,\\
\tilde{\xi}+\tilde{\zeta}=\frac{P \IM S \T \JM}{K^2 S \IM} (\xi+\zeta) \IM ,\\
\tilde{\eta}-\tilde{\zeta}=\frac{P \JM S \T \IM}{K^2 S \JM} (\eta-\zeta) \JM.\\
\end{array}
\end{equation}
Therefore formulae (\ref{DBt7}) constitute the BTs
for the Toda type system (\ref{dyn_triangle}) on the triangular lattice.

We would like to mention that the first two equations of \mref{DBT7} can be easily inverted
\begin{equation}
\begin{array}{l}
\Dim \frac{\Psi}{\T} =\tilde{B}\frac{1}{K \J} \tilde{\Psi} +\tilde{S}\IJ \frac{1}{K \J}\tilde{\Psi} \I 
- (\tilde{B} \frac{1}{K} + \tilde{S}\IJ \frac{1}{K \I} )\Psi \J,\\
\Djm \frac{\Psi}{\T} =
-\tilde{A}\frac{1}{K \I} \tilde{\Psi} -\tilde{S}\IJ \frac{1}{K \I}\tilde{\Psi} \J
+(\tilde{A} \frac{1}{K} + \tilde{S}\IJ \frac{1}{K \J} )\Psi \I.
 \end{array}
\end{equation}
In addition,
equation \mref{K} and the fourth  equation of \mref{DBt7} can be rewritten by means of "new" solutions as follows
\[ \frac{d}{d t}\left(\frac{1}{K}\right)=\frac{1}{2}\left[
\tilde{\xi}\left(\frac{\tilde{A}}{K \I}-\frac{\tilde{A}\IM}{K \IM}\right)+
\tilde{\eta}\left(\frac{\tilde{B}}{K \J}-\frac{\tilde{B}\JM}{K \JM}\right)+
\tilde{\zeta}\left(\frac{\tilde{S} \J}{K \IMJ}-\frac{\tilde{S}\I}{K \IJM}\right)
\right]
\]
\[\frac{\tilde{A}}{K \I}+\frac{\tilde{A}\IM}{K \IM}+
  \frac{\tilde{B}}{K \J}+\frac{\tilde{B}\JM}{K \JM}+
\frac{\tilde{S} \J}{K \IMJ}+\frac{\tilde{S}\I}{K \IJM}=\frac{\tilde{F}}{K}
\]:
i.e., $\frac{1}{K}$ is eigenfunction of  tilded Lax pair.

On the level of the $\tau$-functions, the transformation is given as follows
\beq \tau \mapsto K \check\tau_{-1}, \quad
\hat\tau \mapsto \theta_{-2}\tau_{-2}, \quad
\check\tau \mapsto \left( \frac{\theta \tau\check\tau_{-1} +
\theta_{-1}\tau_{-1}\check\tau + \theta_{-2}\tau_{-2} \check\tau_3}
{\hat\tau_{-1}} \right)_{-2}.
\eeq

As before, the DBTs (\ref{DBt7}) are consistent with all the  reductions of the Toda type system 
(\ref{dyn_triangle}). 

The spatial parts of the above DBTs were already written in \cite{NSD}; the temporal parts, describing the 
time dependence of the transformed solution $\tilde\Psi$, and the transformation 
law for the coefficient $\xi,\eta,\zeta$, are new ingredients of this paper.

\vskip 10pt
\noindent
{\bf Acknowledgements}. This work was supported by the cultural and 
scientific agreements between the University of Roma ``La Sapienza'' and the 
University of Warsaw and the University of Warmia and Mazury in Olsztyn.  
At an initial stage of preparation of the manuscript, M.~N. 
was  supported by European Community under a Marie Curie Intra-European
Fellowship contract no MEIF-CT-2005-011228. A.~D. and M.~N.
were supported by the 
Polish Ministry of Science and Higher Education research grant 1~P03B~017~28.

\end{document}